\documentclass[11pt,a4paper]{article}
\usepackage[a4paper,hmargin=2.8cm,vmargin=3cm]{geometry}
\usepackage[utf8x]{inputenc}
\usepackage[bookmarksnumbered=true]{hyperref}
\usepackage{amsmath,amsthm,amsfonts,amssymb}

\newcommand{\rd}{{\rm d}}
\newcommand{\tr}{{\rm tr}}
\newcommand{\wt}{\widetilde}
\newcommand{\ph}{\varphi}
\newcommand{\PP}{\mathbb{P}}

\newcommand{\cF}{{\cal F}}
\newcommand{\cN}{{\cal N}}
\newcommand{\cL}{{\cal L}}
\newcommand{\cH}{{\cal H}}
\newcommand{\cU}{{\cal U}}

\newcommand{\bR}{{\mathbb R}}
\newcommand{\bC}{{\mathbb C}}
\newcommand{\bN}{{\mathbb N}}

\newtheorem{theorem}{Theorem}[section]

\newtheorem{proposition}[theorem]{Proposition}

\theoremstyle{definition}

\title{Effective equations for quantum dynamics}

\author{Benjamin Schlein \thanks{The author is partially supported by an ERC Starting Grant} \\ \\ Institute of Applied Mathematics, University of Bonn\\ Endenicher Allee 60, 53115 Bonn, Germany}

\begin{document}

\maketitle

\begin{abstract}
We report on recent results concerning the derivation of effective evolution equations starting from many body quantum dynamics. In particular, we obtain rigorous derivations of nonlinear Hartree equations in the bosonic mean field limit, with precise bounds on the rate of convergence. Moreover, we present a central limit theorem for the fluctuations around the Hartree dynamics.
\end{abstract}

\section{Introduction}

We consider systems of $N$ particles, which can be described in quantum mechanics by a wave function $\psi_N \in L^2 (\bR^{3N}, \rd x_1 \dots \rd x_N)$. Here $x_1 \dots , x_N \in \bR^3$ parametrize the positions of the $N$ particles. In nature, one distinguish between bosonic and fermionic systems, whose wave functions are characterized by different permutation symmetries. In these notes, we will focus exclusively on bosonic systems; correspondingly, we will always assume $\psi_N$ to be symmetric with respect to permutations of the $N$ particles. 

\medskip

The wave function $\psi_N$ has a probabilistic interpretation. More precisely, the absolute value $|\psi_N (x_1, \dots, x_N)|^2$ is interpreted as the probability density for finding the $N$ particles close to $x_1, \dots , x_N$. According to this probabilistic interpretation, we will always assume the normalization $\| \psi_N \|_2 = 1$. Apart describing the distribution of the $N$ particles in space, the wave function $\psi_N$ also determines the probability law of every other observable. A physical observable in quantum mechanics is a self-adjoint operator $B$ on $L^2 (\bR^{3N})$. If, for example, $B = \sum_j \lambda_j |\pi_j \rangle \langle \pi_j|$, for a sequence $\{ \lambda_j \}_j$ of real numbers and an orthonormal basis $\{ \pi_j \}_j$ of $L^2 (\bR^{3N})$, the expectation of $B$ in the state described by the wave function $\psi_N$ is given by the $L^2$-product
\[ \langle \psi_N , B \psi_N \rangle = \sum_j \lambda_j \, |\langle \pi_j , \psi_N \rangle |^2 \,. \]
This formula is interpreted as follows; the physical observable $B$ assumes the value $\lambda_j$ with probability $|\langle \pi_j , \psi_N \rangle |^2$. The normalization of the wave function guarantees that
\[ \sum_j |\langle \pi_j , \psi_N \rangle |^2 = \| \psi_N \|_2^2 = 1\,. \]
A similar interpretation can be easily obtained for observables with continuous spectrum. The fact that the same wave function determines the probability law of all physical observables implies that  measurements of non-commuting observables are not independent. This is basis for Heisenberg's uncertainty principle. 

\medskip

The time evolution of an $N$ particle wave function $\psi_N \in L^2 (\bR^{3N})$ is governed by the many-body Schr\"odinger equation 
\begin{equation}\label{eq:schr}  i \partial_t \psi_{N,t} = H_N \psi_{N,t} \, . \end{equation} 
The subscript $t$ indicates the time-dependence of the wave function. On the right hand side of (\ref{eq:schr}), $H_N$ is a self-adjoint operator on the Hilbert space $L^2 (\bR^{3N})$, known as the Hamilton operator of the system. We will consider Hamilton operators of the form 
\[ H_N = \sum_{j=1}^N -\Delta_{x_j} + \lambda \sum_{i<j}^N V(x_i -x_j) \]
where $V$ models the interactions, and $\lambda \in \bR$ is a coupling constant (one could also introduce an external potential; the analysis presented in these notes would still apply). The Schr\"odinger equation (\ref{eq:schr}) is a linear partial differential equation. It always has a unique global solution, which can be obtained by applying the unitary group generated by $H_N$ to the initial wave function $\psi_{N,t=0}$, i.e. $\psi_{N,t} = e^{-iH_N t} \, \psi_{N, 0}$. Hence, the well-posedness of the many-body Schr\"odinger equation is never an issue. What makes the study of the Sch\"odinger equation challenging from the mathematical point of view is the fact that, in typical systems of interest in physics and chemistry, the number of particles $N$ involved in the dynamics is very large, varying from $N \simeq 10^3$ for extremely dilute Bose Einstein condensates, up to values of the order $N \simeq 10^{23}$ for typical samples in chemistry.  For these huge values of $N$, it is in general impossible to extract useful qualitative or quantitative information from (\ref{eq:schr}), going beyond the mere existence and uniqueness of the solution. For this reason, one of the main goals of non-equilibrium statistical mechanics is the derivation of simpler effective evolution equation which provide a good approximation of (\ref{eq:schr}) in the interesting regime. 

\medskip

A simple regime where such effective equations can be derived is the so called mean field limit of quantum mechanics. The mean field regime is characterized by a large number of very weak collisions among the particles. This limit can be realized by choosing  \[  N \gg 1, \quad \lambda \ll 1, \quad \text{with } \quad N \lambda \simeq 1  \quad \text{fixed.} \] The condition $N \gg 1$ guarantees that each particle experiences a large number of collisions; $\lambda \ll 1$ implies, on the other hand, that each collision is very weak. Finally, $N \lambda$ of order one means that the total force produced by the many weak collisions is of order one, and therefore comparable with the inertia of the particle. To study the time evolution in the mean field regime, we can therefore analyze the dynamics generated by the Hamiltonian
\begin{equation}\label{eq:mf-ham} H_N = \sum_{j=1}^N  -\Delta_{x_j}  + \frac{1}{N} \sum_{i<j}^N V (x_i -x_j) \end{equation}
in the limit of large $N$. In particular, we are interested in the evolution of initially factorized states. We assume therefore that, at time $t=0$,
\[ \psi_{N,0} = \ph^{\otimes N} \quad (\text{meaning that } \psi_{N,0} (x_1, \dots , x_N) = \prod_{j=1}^N \ph (x_j) ) \]
for some $\ph \in L^2 (\bR^3)$. Because of the interaction, factorization is not preserved by the time-evolution. Still, considering the mean-field nature of the interaction, we may expect factorization of the evolved many body wave function to be approximately restored in 
the limit of large $N$. In other words, we may expect that, in an appropriate sense,
\begin{equation}\label{eq:conv} \psi_{N,t} \simeq \ph_t^{\otimes N} \end{equation}
as $N\to \infty$. If we assume that this is indeed the case, it is simple to derive a self-consistent equation for the evolution of $\ph_t$. In fact, (approximate) factorization of $\psi_{N,t}$  implies that the $N$ particles are distributed independently in space, with the common probability density function $|\ph_t|^2$. This means that the total potential experienced, say, by particle $j$, can be approximated by
\[ \frac{1}{N} \sum_{i \not = j} V (x_i - x_j) \simeq \frac{1}{N} \sum_{i \neq j} \int \rd x_i \; V(x_i -x_j) |\ph_t (x_i)|^2 \simeq\; (V * |\ph_t|^2) (x_j) \,. \]
Hence, if factorization is preserved in the limit of large $N$, $\ph_t$ must evolve according to the self-consistent nonlinear Hartree equation 
\begin{equation}\label{eq:hartree0} i\partial_t \ph_t = -\Delta \ph_t + \, (V * |\ph_t|^2) \ph_t \,\end{equation}
where the many-body interactions have been replaced by an average, mean field, potential $(V * |\ph_t|^2)$. 

\medskip

In which sense can we expect (\ref{eq:conv}) to hold true? To answer this question, we introduce the reduced density matrices associated with the solution of the Schr\"odinger equation $\psi_{N,t}$. For $k=1,2, \dots , N$, we define the $k$-particle reduced density as
\[ \gamma^{(k)}_{N,t} = \tr_{k+1, \dots , N} \; |\psi_{N,t} \rangle \langle \psi_{N,t}| \]
where $\tr_{k+1, \dots , N}$ denotes the partial trace over the last $(N-k)$ particles, and where $|\psi_{N,t} \rangle \langle \psi_{N,t}|$ denotes the orthogonal projection onto $\psi_{N,t}$. In other words, $\gamma^{(k)}_{N,t}$ is defined as the non-negative trace class operator on $L^2 (\bR^{3k})$, with the kernel 
\[\begin{split}  &\gamma^{(k)}_{N,t} (x_1, \dots , x_k ; x'_1, \dots , x'_k) \\ &= \int \rd x_{k+1} \dots \rd x_{N}  \;
\psi_{N,t} (x_1,\dots , x_k, x_{k+1}, \dots , x_{N}) \overline{\psi}_{N,t} (x'_1, \dots , x'_k, x_{k+1}, \dots , x_N) . \end{split} \]
{F}rom the normalization of $\psi_{N,t}$, we find $\tr \; \gamma_{N,t}^{(k)} =1$ for all $k,t,N$. Of course, for $k < N$, the $k$-particle density $\gamma^{(k)}_{N,t}$ does not contain the full information about the state described by $\psi_{N,t}$. Nevertheless, $\gamma^{(k)}_{N,t}$ is sufficient to compute the expectation of an arbitrary $k$-particle observable. In other words, 
\[ \left\langle \psi_{N,t} , \left( J^{(k)} \otimes 1^{(N-k)} \right) \psi_{N,t} \right\rangle = \tr \; \left( J^{(k)} \otimes 1^{(N-k)} \right) \, |\psi_{N,t} \rangle \langle \psi_{N,t} | = \tr \; J^{(k)} \, \gamma^{(k)}_{N,t} \]
for any self-adjoint operator $J^{(k)}$ on $L^2 (\bR^{3k})$. It turns out that reduced density matrices provide the right language to describe the convergence of the full many-body Schr\"odinger evolution generated by (\ref{eq:mf-ham}) towards the limiting mean field Hartree dynamics (\ref{eq:hartree0}). Under appropriate assumptions on the potentials, we have the following theorem. 
\begin{theorem}\label{thm:mf}
Assume that, at time $t=0$, $\psi_{N} = \ph^{\otimes N}$, for $\ph \in H^1 (\bR^3)$. Let $\psi_{N,t} = e^{-iH_N t} \psi_N$ be the solution of the Schr\"odinger equation with mean field Hamiltonian (\ref{eq:mf-ham}), and with initial data $\psi_N$. Then, for every fixed $k \in \bN$, $t \in \bR$, we have
\[ \gamma^{(k)}_{N,t} \to |\ph_t \rangle \langle \ph_t|^{\otimes k} \]
as $N \to \infty$, with respect to the trace class topology. Here $\ph_t$ is the solution of the nonlinear Hartree equation (\ref{eq:hartree0}), with initial data $\ph_{t=0} = \ph$. 
\end{theorem}
This theorem explains in which sense (\ref{eq:conv}) should be understood; the reduced densities of $\psi_{N,t}$ converge, as $N \to \infty$, towards the reduced densities of the product $\ph_t^{\otimes N}$. In terms of expectation of observables, the theorem implies that, for any bounded self-adjoint operator $J^{(k)}$ on $L^2 (\bR^{3k})$,
\[ \left\langle \psi_{N,t} , \left( J^{(k)} \otimes 1^{(N-k)} \right) \psi_{N,t} \right\rangle \to \langle \ph_t^{\otimes k} , J^{(k)} \ph_t^{\otimes k} \rangle \]
as $N \to \infty$. In other words, if we compute the expectation of an observable depending only on a fixed number of particles $k$ then, in the limit of large $N$, we can replace the solution of the full many-body Schr\"odinger equation with products of the solution of the one-particle Hartree equation. 

\medskip

The first proof of Theorem \ref{thm:mf} was obtained in \cite{S} for the case of bounded potentials $\| V \|_\infty < \infty$. The techniques introduced in \cite{S} were then extended in \cite{EY} to potentials with a Coulomb singularity $V(x) = \pm 1/ |x|$. In \cite{RS}, a different approach was then developed to show the convergence towards the Hartree dynamics for potentials with Coulomb singularities. In contrast with the previous methods, this new approach provided precise bounds on the rate of the convergence towards the limiting Hartree dynamics. More recently, a different derivation, covering also potentials with more severe singularities, was introduced in \cite{KP}. A more complete list of related works 
can be found in \cite{CLS,BKS}. In the rest of the paper, the approach developed in \cite{RS} will be presented, and some of its applications will be discussed.

\section{Fock space representation and evolution of coherent states}

We define the bosonic Fock space over $L^2 (\bR^3)$ as
\[ \cF = \bC \oplus \bigoplus_{n \geq 1} L^2_s (\bR^{3n} , \rd x_1 \dots \rd x_n) \, ,\]
where $L^2_s (\bR^{3n})$ denotes the subspace of $L^2 (\bR^{3n})$ consisting of permutation symmetric functions (in accordance with the bosonic symmetry). Elements of $\cF$ are sequences 
$\psi = \{ \psi^{(n)} \}_{n \geq 1}$, where $\psi^{(n)} \in L^2_s (\bR^{3n})$ is an $n$-boson wave function. $\cF$ is a Hilbert space with respect to the inner product 
\[ \langle \psi ; \ph \rangle = \overline{\psi^{(0)}} \ph^{(0)} +
\sum_{n \geq 1} \langle \psi^{(n)}, \ph^{(n)} \rangle \, .\]
The idea beyond the construction of the Fock space is that, on $\cF$, one can describe states with variable number of particles. The state described by the sequence $\psi = \{ \psi^{(n)} \}_{n \geq 0}$ has $n$ particles with probability $\| \psi^{(n)} \|_2^2$ (the normalization $\| \psi \|_\cF = 1$ is always imposed). 

\medskip

To measure the number of particles in a state $\psi$, we introduce the number of particles operator $\cN$ defined by 
\[ (\cN \psi)^{(n)} = n \psi^{(n)}\,. \]
Eigenvectors of $\cN$ are sequences of the form $\{ 0, 0, \dots , 0, \psi^{(n)}, 0, \dots \}$ having only one non-zero component. A special example of such a state is the vacuum $\Omega = \{ 1, 0 , \dots \}$,   describing a state with no particles at all. 

\medskip

In order to define a time-evolution on $\cF$, we introduce the Hamilton operator $\cH_N$, defined by
\begin{equation}\label{eq:ham-F1}
\begin{split} 
(\cH_N \psi)^{(n)} &= \cH_N^{(n)} \psi^{(n)}  \\
\cH_N^{(n)} &= \sum_{j=1}^n -\Delta_{x_j} + \frac{1}{N} \sum_{i<j}^n V (x_i -x_j)  \,.
\end{split} \end{equation}
By definition, the Hamilton operator $\cH_N$ leaves all sectors with fixed number of particles invariant. In particular, when applied on states with exactly $N$ particles, the Hamiltonian $\cH_N$ acts exactly as the mean field Hamiltonian (\ref{eq:mf-ham}). 

\medskip

On $\cF$ it is useful to introduce creation and annihilation operators. For $f \in L^2 (\bR^3)$, we define namely the creation operator $a^* (f)$ and the annihilation operator $a(f)$ by the formulas
\begin{equation*}\begin{split} \left( a^* (f) \psi \right)^{(n)}
(x_1, \dots ,x_n) &= \frac{1}{\sqrt{n}} \sum_{j=1}^n f (x_j)
\psi^{(n-1)} (x_1 , \dots, \hat{x}_j,\dots, x_n)
\\ \left( a (f) \psi
\right)^{(n)} (x_1, \dots ,x_n) &= \sqrt{n+1} \int \rd x \;
\overline{f (x)} \psi^{(n+1)} (x, x_1 , \dots , x_n) \,. \end{split}
\end{equation*}
It is simple to check that creation and annihilation operators are adjoint to each other. Also, they satisfy the canonical commutation relations 
\begin{equation}\label{eq:CCR} [a(f), a^* (g)] = (f,g)_{L^2}
\quad [ a(f), a(g)] = [a^* (f) , a^* (g)] = 0 \,.\end{equation}
Applying creation operators on the vacuum, it is possible to generate an orthonormal basis for the Hilbert space $\cF$. For example, the $N$-particle state $\ph^{\otimes N}$ can be expressed as
\[ \{ 0 ,  \dots , 0 , \ph^{\otimes N}, 0, \dots \} =  \frac{1}{\sqrt{N!}} (a^* (\ph))^N \Omega \, . \]
We will also make use of operator valued distributions $a_x^*, a_x$, defined by 
\[ a^* (f) = \int dx \, f (x) a^*_x \qquad  a(f) =
\int \rd x \; \overline{f(x)} \, a_x \,. \]

\medskip

In terms of these operator valued distributions, we can express the number of particles operator as
\[ \cN = \int dx \, a^*_x a_x \,. \]
A simple consequence of this formula is that, although creation and annihilation operators are not bounded, they are nevertheless bounded with respect to the square root of the number of particle operator, in the sense that
\begin{equation}\label{eq:bd-aaN}
\| a (f) \psi \| \leq \| f \|_2 \| \cN^{1/2} \psi \|  \quad \text{and } \quad 
\| a^* (f) \psi \| \leq \|f \|_2 \| (\cN+1)^{1/2} \psi \|
\end{equation}
for all $f \in L^2 (\bR^3)$, $\psi \in \cF$.  

\medskip

Also the Hamilton operator $\cH_N$ can be expressed in terms of the operator valued distributions $a_x^*, a_x$ as 
\begin{equation}\label{eq:ham-F2} \cH_N = \int dx \, \nabla_x a^*_x \nabla_x a_x + \frac{1}{N} \int dx dy V(x-y) a_x^* a_y^* a_y a_x \, . \end{equation}
The fact that the Hamiltonian commutes with $\cN$ is reflected, in (\ref{eq:ham-F2}), by the observation  that each term in $\cH_N$ contains the same number of creation and annihilation operators. As a consequence the evolution generated by $\cH_N$ leaves the sectors of $\cF$ with fixed number 
of particles invariant. In particular, if we consider the evolution of an initially factorized state, we find 
\[ e^{-i\cH_N t} \{ 0,\dots, 0 , \ph^{\otimes N} , 0 ,\dots \} = \{ 0, \dots, 0 , e^{-iH_N t} \ph^{\otimes N} , 0 ,\dots \} \]
where $H_N$ is the mean field Hamiltonian defined in (\ref{eq:mf-ham}). What did we gain then by switching to a Fock space representation of the many boson system? The gain is that now we have much more freedom in the choice of the initial state. We will make use of this freedom by considering so called coherent states as initial data. 

\medskip

For $f \in L^2 (\bR^3)$, we define the Weyl operator 
\[ W(f) = \exp \left(a^* (f) - a(f) \right) \, . \]
The coherent state with wave function $f$ is then defined as 
\begin{equation}\label{eq:coh} \begin{split} W(f) \Omega & = e^{-\| f \|^2/2} \sum_{n \geq 0}  \frac{(a^* (f))^n}{n!} \Omega = e^{-\| f \|^2/2} \left\{ 1 , f , \frac{f^{\otimes 2}}{\sqrt{2!}} , \dots \right\} \, .\end{split}
\end{equation}
Since $W(f)$ is a unitary operator on $\cF$, coherent states are always normalized. It is clear from  (\ref{eq:coh}) that coherent states do not have a fixed number of particles; instead they are given by a linear combination of states with all possible number of particles. Still, one can ask what is the expected number of particles in the coherent state $W(f) \Omega$. A simple computation shows that
\[ \left\langle W(f) \Omega, \cN W(f)\Omega \right\rangle = \|\ph \|^2 . \]
This follows from the fact that Weyl operators act on creation and annihilation operators by simple shifts, in the sense that
\begin{equation}\label{eq:Wshift}
\begin{split}
W^* (f) \, a^* (g) \, W(f) &= a^* (g) + \langle f , g \rangle \\
W^* (f) \, a (g) \, W(f) &= a (g) + \langle g , f  \rangle\,.
\end{split}
\end{equation}
The property (\ref{eq:Wshift}) also implies that coherent states are eigenvectors of all annihilation operators, i.e.:
\[ a(g) W(f) \Omega = \langle g ,f \rangle W(f) \Omega \]
for every $f,g\in L^2 (\bR^3)$. This important observation is the main reason for the nice algebraic properties of coherent states which, as we will see below, significantly simplify the analysis of their time evolution.

\medskip

We study next the evolution of coherent states. In order to recover the mean field regime discussed in the introduction, the number of particles must be related with the parameter $N$ appearing in the Hamiltonian (\ref{eq:ham-F1}) (equivalent to (\ref{eq:ham-F2})). Since we want the initial state to be coherent, we cannot fix the number of particles to be equal to $N$. Still, we can require $N$ to be the expected number of particles. Hence, we fix $\ph \in L^2 (\bR^3)$ with $\| \ph \|_2 =1$, and consider the initial coherent state \[ \Psi_N = W(\sqrt{N} \ph) \Omega\, .  \]
We study its time evolution \[ \Psi_{N,t} = e^{-i \cH_N t} \Psi_N = e^{-i \cH_N t} W(\sqrt{N} \ph) \Omega \]
as generated by (\ref{eq:ham-F2}). In particular, we would like to compute the reduced densities $\Gamma_{N,t}^{(k)}$ associated with $\Psi_{N,t}$. Let us consider for example the one-particle reduced density. It turns out that the kernel $\Gamma_{N,t}^{(1)} (x,y)$ of the one-particle reduced density can be expressed as  
\begin{equation}\label{eq:exp1}\begin{split}  \Gamma_{N,t}^{(1)} (x,y) = \; & \frac{1}{N} \left\langle \Psi_{N,t} , a_y^* a_x \Psi_{N,t} \right\rangle  \\ =  \; & \frac{1}{N} \left\langle e^{-i\cH_N t} W
(\sqrt{N} \ph) \Omega, \, a_y^* a_x \, e^{-i\cH_N t} W (\sqrt{N}
\ph) \Omega \right\rangle \, .
\end{split} \end{equation}
Of course, for $t \not = 0$, $\Psi_{N,t}$ is not a coherent state. Nevertheless, because of the mean-field character of the interaction, in the limit of large $N$ we can expect it to be approximately coherent, with an evolved one-particle wave function $\sqrt{N} \ph_t$, where $\ph_t$ solves the Hartree equation (\ref{eq:hartree0}). In other words, we may expect that \[ \Psi_{N,t} = e^{-iH_N t} W (\sqrt{N} \ph) \Omega \simeq W (\sqrt{N} \ph_t) \Omega \, . \]
Recalling that coherent states are eigenvectors of the annihilation operators, it seems appropriate to expand the operator valued distributions $a_x, a_y^*$ on the right hand side of (\ref{eq:exp1}) around their mean field values $\sqrt{N} \ph_t (x)$ and $\sqrt{N} \overline{\ph}_t (y)$. Writing $a_x = \sqrt{N} \ph_t (x) + (a_x - \sqrt{N} \ph_t (x))$ and similarly for $a^*_y$, and inserting in (\ref{eq:exp1}), we find
\begin{equation}\label{eq:exp2} \begin{split}
\Gamma^{(1)}_{N,t} &(x;y) - \ph_t (x) \overline{\ph}_t (y)
\\ = \; &
\frac{1}{N} \left\langle \Omega, W^* (\sqrt{N} \ph) e^{i\cH_N t} \;
(a_y^* - \sqrt{N} \overline{\ph}_t (y))( a_x - \sqrt{N} \ph_t (x)) \, e^{-i\cH_N t} W
(\sqrt{N} \ph) \Omega \right\rangle
\\ &+ \frac{\ph_t (x)}{\sqrt{N}}\left\langle \Omega, W^*
(\sqrt{N} \ph) e^{i\cH_N t} \; \left(a_y^* - \sqrt{N}
\overline{\ph}_t (y) \right) \,
e^{-i\cH_N t} W (\sqrt{N} \ph) \Omega \right\rangle \\
&+ \frac{\overline{\ph}_t (y)}{\sqrt{N}} \left\langle \Omega, W^*
(\sqrt{N} \ph) e^{i\cH_N t} \; \left(a_x - \sqrt{N} \ph_t (x)\right)
\, e^{-i\cH_N t} W (\sqrt{N} \ph) \Omega \right\rangle \, .
\end{split}
\end{equation}
Observe here that $\ph_t (x) \overline{\ph}_t (y)$ is the kernel of the orthogonal projection $|\ph_t\rangle \langle \ph_t|$ onto the solution of the Hartree equation (\ref{eq:hartree0}). To establish the convergence of the full many-body evolution towards the limiting Hartree dynamics, we need therefore to bound the error terms on the r.h.s. of (\ref{eq:exp2}). 

\medskip

Next, we make use of an idea introduced in the related context of the classical limit of quantum mechanics in \cite{H}. We observe namely that, from (\ref{eq:Wshift}), the fluctuations of the operator valued distributions $a_y^* ,a_x$ around their mean field values can be expressed as 
\[ \begin{split} (a_y^* - \sqrt{N}\, \overline{\ph}_t (y)) &= W (\sqrt{N} \ph_t) \,  a_y^* \, W^* (\sqrt{N} \ph_t) \\ (a_x - \sqrt{N} \, \ph_t (y)) &= W (\sqrt{N} \ph_t ) \,  a_x \, W^* (\sqrt{N} \ph_t) \, . \end{split}\]
{F}rom (\ref{eq:exp2}), we conclude that 
\begin{equation*} \begin{split}
\Gamma^{(1)}_{N,t} &(x;y) - \ph_t (x) \overline{\ph}_t (y)
\\ = \; &
\frac{1}{N} \left\langle \Omega, W^* (\sqrt{N} \ph) e^{i\cH_N t} W(\sqrt{N} \ph_t) \;
a_y^* a_x W^* (\sqrt{N} \ph_t) \, e^{-i\cH_N t} W
(\sqrt{N} \ph) \Omega \right\rangle
\\ &+ \frac{\ph_t (x)}{\sqrt{N}} \left\langle \Omega, W^*
(\sqrt{N} \ph) e^{i\cH_N t} \; W (\sqrt{N} \ph_t) a_y^* W^* (\sqrt{N} \ph_t) 
e^{-i\cH_N t} W (\sqrt{N} \ph) \Omega \right\rangle \\
&+ \frac{\overline{\ph}_t (y)}{\sqrt{N}} \left\langle \Omega, W^*
(\sqrt{N} \ph) e^{i\cH_N t} W (\sqrt{N} \ph_t) a_x 
W^* (\sqrt{N} \ph_t) \, e^{-i\cH_N t} W (\sqrt{N} \ph) \Omega \right\rangle \, .
\end{split}
\end{equation*}
Introducing the fluctuation dynamics described by the two-parameter group of unitary transformations 
\begin{equation}\label{eq:fluc} \cU (t;s) = W (\sqrt{N} \ph_t) e^{-i \cH_N (t-s)} W^* (\sqrt{N} \ph_s), \qquad \text{for all $t,s \in \bR$,} \end{equation}
we obtain the identity
\begin{equation}\label{eq:G-ph} \begin{split}
\Gamma^{(1)}_{N,t} &(x;y) - \ph_t  (x) \overline{\ph}_t (y) \\ = \; &\frac{1}{N} \left\langle \Omega, \cU^* (t;0) \, a_y^* a_x  \, \cU (t;0) \Omega \right\rangle
\\ & + \frac{\ph_t (x)}{\sqrt{N}} \left\langle \Omega, \cU^* (t;0) \, a_y^* \, \cU (t;0) \Omega \right\rangle + \frac{\overline{\ph}_t (y)}{\sqrt{N}} \left\langle \Omega, \cU^* (t;0) \, a_x \, \cU(t;0)  \Omega \right\rangle \, .
\end{split}
\end{equation}
Let us focus on the first error term on the r.h.s. of (\ref{eq:G-ph}). Since, by  (\ref{eq:bd-aaN}), creation and annihilation operators are bounded with respect to the square root of the number of particles operator, the smallness of this term follows if we can show uniform (in $N$) estimates for the growth of the expectation of $\cN$ with respect to the dynamics $\cU (t;0)$. To obtain such bounds, we observe that the fluctuation dynamics satisfies 
\[ i \partial_t \cU (t;s) = \cL_N (t) \cU (t;s) \quad 
\text{with } \quad \cU (s;s) = 1 \quad \text{for all $s \in \bR$} \]
with the time-dependent generator
\begin{equation}\begin{split} 
\label{eq:LNt} \cL_N (t)  = &\; \int \rd x \; \nabla_x a^*_x \nabla_x a_x + \int dx \, (V * |\ph_t|^2) (x) a_x^* a_x
\\ &+ \int \rd x \rd y \; V (x-y) \ph_t (x) \overline{\ph}_t (y) \,
a_x^* a_y \\ &+ \int \rd x \rd y \; V (x-y) \left(\ph_t (x) \ph_t
(y) \, a_x^* a^*_y + \overline{\ph}_t (x) \overline{\ph}_t (y) a_x
a_y \right) \\ &+\frac{1}{\sqrt{N}} \int \rd x \rd y \, V(x-y) \,
a_x^*
\left( \overline{\ph}_t (y) a_y + \ph_t (y) a_y^* \right) a_x \\
&+\frac{1}{N} \int \rd x \rd y \, V(x-y) \, a^*_x a^*_y a_y a_x \, .
\end{split}
\end{equation}
In contrast with $\cH_N$, the generator $\cL_N (t)$ (in particular, the terms on the third and fourth line) does not commute with $\cN$. As a consequence, the fluctuation dynamics $\cU (t;s)$ does not preserve the number of particles. This is hardly surprising, since fluctuations around the mean field solution have to grow during the time evolution. Although the expectation of the number of particles is not preserved under $\cU (t;s)$, it is possible to show uniform bounds for the growth of $\cN$ and any of its power. The following result was proven in \cite{RS}. 
\begin{proposition} \label{thm:flu-bd}
Suppose that there exists a constant $D >0$ such that the operator inequality 
$V^2 (x) \leq D (1-\Delta)$ holds true. Suppose moreover that $\cU (t;s)$ is defined as in (\ref{eq:fluc}). Then, for every $k \in \bN$ there exist constants $C,K >0$ with 
\[ \left\langle \psi , \cU^* (t;s) \, (\cN+1)^k \, \cU (t;s) \psi \right\rangle \leq C e^{K |t-s|} \langle \psi, (\cN+1)^{2k+2} \, \psi \rangle \]
for all $t,s \in \bR$.
\end{proposition}
{\it Remark.} The operator inequality $V^2 (x) \leq D (1-\Delta)$, meaning that
\[ \int dx \, V(x) \, |\ph (x)|^2 \leq D \int dx \, \left[ |\nabla \ph (x)|^2 + |\ph (x)|^2 \right] = D \| \ph \|_{H^1}^2 \, , \] for all $\ph \in L^2 (\bR^3)$,  is satisfied, because of Hardy's inequality, for potentials with Coulomb type singularities $V(x) = \pm 1/ |x|$. 

\medskip

Proposition \ref{thm:flu-bd} immediately implies that the first error term on the r.h.s. of (\ref{eq:G-ph}) is of the order $1/N$, for any fixed $t \in \bR$. With some more work one can show the same estimate also for the last two terms on the r.h.s. of (\ref{eq:G-ph}). As a consequence, one obtains convergence towards the Hartree dynamics. The details of the proof of the next theorem can be found in \cite{RS}.
\begin{theorem}
Suppose that there exists a constant $D >0$ such that the operator inequality 
$V^2 (x) \leq D (1-\Delta)$ holds true. Suppose that, at time $t = 0$, $\Psi_N = W(\sqrt{N} \ph) \Omega$, for some $\ph \in H^1 (\bR^3)$. Let $\Psi_{N,t} = e^{-i \cH_N t} \Psi_N$, with the Hamilton operator (\ref{eq:ham-F2}), and let $\Gamma_{N,t}^{(1)}$ be the one-particle density associated with $\Psi_{N,t}$. Then there exist constants $C,K >0$ with
\[ \tr \, \left| \Gamma^{(1)}_{N,t} - |\ph_t \rangle \langle \ph_t | \right| \leq \frac{C e^{K|t|}}{N} \]
where $\ph_t$ is the solution of the Hartree equation (\ref{eq:hartree0}), with initial data $\ph_{t=0} = \ph$. Similar bounds can also be proven for the $k$-particle reduced density, for any fixed $k \in \bN$. 
\end{theorem}
{\it Remark.}  The same result can be obtained, with exactly the same techniques, for initial states of the form $\psi_N = W(\sqrt{N} \ph) \psi$, for a $\psi \in \cF$ with $\langle \psi, \cN^2 \psi \rangle \lesssim 1$ (independent of $N$). 

\medskip

This approach to the study of the evolution of coherent states not only implies the convergence towards the mean field Hartree dynamics. Instead, it also establishes the form of the fluctuation dynamics in the limit of large $N$. Formally, the generator $\cL_N (t)$ converges, as $N \to \infty$, towards the Fock-space operator 
\[ \begin{split} 
\cL_\infty (t) 
= &\; \int \rd x \; \nabla_x a^*_x \nabla_x a_x + \int dx \, (V * |\ph_t|^2) (x) a_x^* a_x
\\ &+ \int \rd x \rd y \; V (x-y) \ph_t (x) \overline{\ph}_t (y) \,
a_x^* a_y \\ &+ \int \rd x \rd y \; V (x-y) \left(\ph_t (x) \ph_t
(y) \, a_x^* a^*_y + \overline{\ph}_t (x) \overline{\ph}_t (y) a_x
a_y \right) \, . \end{split} \]
One can expect, therefore, that the fluctuation dynamics $\cU (t;s)$ converges, as $N \to \infty$, towards the limiting dynamics $\cU_\infty (t;s)$, defined by 
\[ i\partial_t \cU_\infty (t;s) = \cL_\infty (t) \cU_\infty (t;s) \qquad \text{with } \quad \cU_\infty (s;s)  = 1 \quad \text{for all $s \in \bR$} \, . \]
Since $\cL_\infty (t)$ is a time-dependent unbounded operator, the definition of the limiting dynamics $\cU_\infty (t;s)$ generated by $\cL_\infty (t)$ is not at all trivial. The existence of $\cU_\infty (t;s)$, and the (strong) convergence $\cU (t;s) \to \cU_\infty (t;s)$ were rigorously established in \cite{GV}, making use of appropriate approximations of $\cL_\infty (t)$.

\medskip

Since the generator $\cL_\infty (t)$ is a quadratic expression in creation and annihilation operators, it turns out that the limiting dynamics $\cU_\infty (t;s)$ can be described as a so called Bogoliubov transformation.  For $f,g \in L^2 (\bR^3)$, we define on $\cF$ the linear combination of creation and annihilation operators 
\[ A(f,g) = a^* (f) + a (\overline{g}) .\]
By definition, $A$ is linear in both $f$ and $g$. Observe that
\[ (A(f,g))^* = A(Jg, Jf) = A \left(  \left(\begin{array}{ll} 0 & J \\ J & 0 \end{array} \right) \left( \begin{array}{l} f \\ g \end{array} \right) \right) \]
where $J$ is the antiunitary operator on $L^2 (\bR^3)$ defined by $Jf = \overline{f}$. A simple computation shows that, in terms of the operators $A(f,g)$, the canonical commutation relations (\ref{eq:CCR}) assume the form
\[ \left[ A(f_1,g_1) , A(f_2 , g_2) \right] = \left \langle \left( \begin{array}{l} f_1 \\ g_1 \end{array} \right) , 
\left(\begin{array}{ll} 1 & 0 \\  0 & -1  \end{array} \right)  \left( \begin{array}{l} f_2 \\ g_2 \end{array} \right) \right\rangle \]
where $\langle . , . \rangle$ denotes the standard inner product on $L^2 (\bR^3) \oplus L^2 (\bR^3)$. 
A Bogoliubov transformation is a linear map $\theta : L^2 (\bR^3) \oplus L^2 (\bR^3) \to L^2 (\bR^3) \oplus L^2 (\bR^3)$ with the properties
\begin{equation}\label{eq:conj} \theta \left(\begin{array}{ll} 0 & J \\ J & 0 \end{array} \right) = \left(\begin{array}{ll} 0 & J \\ J & 0 \end{array} \right)  \theta \end{equation}
and \begin{equation}\label{eq:CCR-BT} \theta^* \left(\begin{array}{ll} 1 & 0 \\  0 & -1  \end{array} \right) \theta = \left(\begin{array}{ll} 1 & 0 \\  0 & -1  \end{array} \right) \, . \end{equation}
Eq. (\ref{eq:conj}) guarantees the preservation of the relation between $A$ and its adjoint. Eq. (\ref{eq:CCR-BT}), on the other hand, guarantees the preservation of the canonical commutation relations. It is simple to check that every Bogoliubov transformation can be expressed through the operator-valued matrix
\begin{equation}\label{eq:UV} \theta = \left(\begin{array}{ll} U & J V J \\ V & J U J \end{array} \right) \end{equation}
where $U,V : L^2 (\bR^3) \to L^2 (\bR^3)$ are s.t. $U^* U - V^* V = 1$ and $U^* \overline{V} - V^* \overline{U} = 0$ (notice that $\theta$ is not unitary, unless $V = 0$). Every Bogoliubov transformation defines therefore a new set of creation and annihilation operators. It is interesting to ask when the new representation of the canonical commutation relation is unitary equivalent to the one given by the original operators $a^* (f), a (f)$. It turns out that this is the case if and only if the operator $V$ appearing in (\ref{eq:UV}) is Hilbert-Schmidt (Shale-Stinespring condition). 

\medskip

The statement that the limiting fluctuation dynamics $\cU_\infty (t;s)$ acts as a Bogoliubov transformation has to be understood as follows. There exists a two parameter family of Bogoliubov transformation $\theta (t;s) : L^2 (\bR^3) \oplus L^2 (\bR^3) \to L^2 (\bR^3) \oplus L^2 (\bR^3)$ such that 
\begin{equation}\label{eq:BTU} \cU_\infty^* (t;s) A (f,g) \cU_\infty (t;s) = A (\theta (t;s) (f,g)) \end{equation}
for every $f,g \in L^2 (\bR^3)$.  Formally, $\theta (t;s)$ is given by the solution of 
\[ i \partial_t \theta (t;s) = \left( \begin{array}{ll} D_t & -\overline{B}_t \\ B_t & -\overline{D}_t \end{array} \right) \theta (t;s) \]
with initial condition $\theta (s;s) = 1$, and with $D_t , B_t : L^2 (\bR^3) \to L^2 (\bR^3)$ defined by
\[ \begin{split} D_t  f &= -\Delta f + (V*|\ph_t|^2) f + (V * \overline{\ph}_t f ) \ph_t \\ B_t f &= (V* \overline{\ph}_t f) \overline{\ph}_t \, . \end{split} \]
We will not make use of this formal characterization of $\theta (t;s)$. The important observation is that the time evolution $\cU_\infty (t;s)$, which is in principle a two-parameter family of unitary transformation on the large Hilbert space $\cF$ can be completely described in terms of the family $\theta (t;s)$ operating on the much smaller space $L^2 (\bR^3) \oplus L^2 (\bR^3)$. 

\medskip

So far we considered the evolution of initial coherent states. Next, we turn our attention back to initially factorized (or approximately factorized) $N$-particle states, as those considered in the introduction. To this end, we notice that, for any $\ph \in L^2 (\bR^3)$, we can write the factorized state 
\[ \{ 0, \dots, 0, \ph^{\otimes N}, 0 , \dots \}  = d_N P_N W(\sqrt{N} \ph) \Omega \]
where $P_N$ denotes the orthogonal projection onto the sector with exactly $N$ particles, and where $d_N$ is a normalization constant; a simple computation shows that $d_N \simeq N^{1/4}$. The reduced one-particle density associated with the evolution of the factorized initial data is given by
\[ \begin{split} \gamma^{(1)}_{N,t} (x,y) =\; & \frac{1}{N} \left\langle e^{-i\cH_N t} \frac{(a^* (\ph))^N}{\sqrt{N!}} \Omega, a_x^* a_y  e^{-i\cH_N t} \frac{(a^* (\ph))^N}{\sqrt{N!}} \Omega \right \rangle \\
= \; & \frac{d_N}{N}  \left\langle e^{-i\cH_N t} \frac{(a^* (\ph))^N}{\sqrt{N!}} \Omega, a_x^* a_y  e^{-i\cH_N t} P_N W(\sqrt{N} \ph) \Omega \right \rangle \\
= \; &  \frac{d_N}{N}  \left\langle e^{-i\cH_N t} \frac{(a^* (\ph))^N}{\sqrt{N!}} \Omega, a_x^* a_y  e^{-i\cH_N t} W(\sqrt{N} \ph) \Omega \right \rangle \end{split} \]
because $P_N$ commutes with $\cH_N$ and with $a_y^* a_x$, and because $P_N (a^* (\ph))^N \Omega = (a^* (\ph))^N \Omega$. Letting 
\[ \xi = d_N W^* (\sqrt{N} \ph) \frac{(a^* (\ph))^N}{\sqrt{N!}} \Omega \]
we find
\begin{equation}\label{eq:fact-id} \begin{split}
\gamma^{(1)}_{N,t} (x,y) 
= \; &  \frac{1}{N}  \Big\langle \xi, 
\cU (t;0) (a_x^* + \sqrt{N} \overline{\ph}_t (x)) (a_y + \sqrt{N} \ph_t (y)) \cU^* (t;0) \Omega \Big\rangle \\
= \; &  \overline{\ph}_t (x) \ph_t (y) + \frac{1}{N}  \left\langle \xi, \cU^* (t;0) a^*_x a_y \cU (t;0) \Omega \right \rangle \\ &+
\frac{\overline{\ph}_t (x)}{\sqrt{N}}  \left\langle \xi, \cU^* (t;0) a_y \cU (t;0) \Omega \right \rangle +  \frac{\ph_t (y)}{\sqrt{N}}  \left\langle \xi , \cU^* (t;0) a^*_x \cU (t;0) \Omega \right \rangle \\
\end{split} \end{equation}
where $\cU (t;s)$ denotes the fluctuation dynamics defined in (\ref{eq:fluc}). {F}rom this formula, we conclude that, similarly as for initial coherent states, proving the convergence towards the Hartree dynamics reduces to the problem of obtaining uniform bounds for the growth of the product
\begin{equation}\label{eq:xiN} \left| \left\langle \xi , \cU^* (t;0) \cN \cU (t;0) \Omega \right\rangle \right| \, . \end{equation}
The only difference compared to the case of coherent initial data is that now, on the l.h.s. of the product, we have the vector $\xi$ instead of the vacuum. Since $\| \xi \| = d_N \simeq N^{1/4}$, it seems now more difficult to get estimates uniformly in $N$. It turns out, however, that when restricted to sectors with small number of particles, $\xi$ is an order one vector, in the sense that
\[ \| (\cN + 1)^{-1} \xi \| \lesssim 1 \]
uniformly in $N$. For this reason,
\begin{equation}
\label{eq:bd-fact} \begin{split} 
\left| \left\langle \xi , \cU^* (t;0) \cN \cU (t;0) \Omega \right\rangle \right| \leq \; & \| (\cN+1)^{-1} \xi \| \, \| (\cN+1) \cU^* (t;0) \cN \cU (t;0) \Omega \| \\ \lesssim \; & e^{K |t|} \| (\cN +1)^3 \cU (t;0) \Omega \| \\ \lesssim 
\; & e^{2K |t|} \| (\cN +1)^{7} \Omega \|  \lesssim e^{2K |t|}
\end{split}
\end{equation}
applying twice Proposition \ref{thm:flu-bd}. As a corollary of (\ref{eq:bd-fact}), with some additional work needed to bound the last two terms on the r.h.s. of (\ref{eq:fact-id}), we obtain the convergence of the evolution of initially factorized data towards the Hartree dynamics, with an explicit bound on the rate of convergence. The proof of the following theorem was found in \cite{CLS}, optimizing ideas from \cite{RS} (in \cite{RS}, the error for factorized initial data was shown to be at most of the order $N^{-1/2}$, for every fixed $t \in \bR$). 
\begin{theorem}\label{thm:fact}
Suppose that there exists a constant $D >0$ such that the operator inequality 
$V^2 (x) \leq D (1-\Delta)$ holds true. Let $\psi_{N,t} = e^{-iH_N t} \ph^{\otimes N}$ with the Hamilton operator (\ref{eq:mf-ham}), and for some $\ph \in H^1 (\bR^3)$. Let $\gamma^{(1)}_{N,t}$ be the one-particle reduced density associated with $\psi_{N,t}$. Then there exist constants $C,K >0$ with
\[ \tr \, \left| \gamma^{(1)}_{N,t} - |\ph_t \rangle \langle \ph_t| \right| \leq \frac{Ce^{K |t|}}{N} \, .\]
Similar bounds holds for the $k$-particle reduced densities as well.
\end{theorem}
{\it Remark.}  The same result can be obtained for initial data of the form $\psi_N = d_N P_N W(\sqrt{N} \ph) \psi$, for an arbitrary $\psi \in \cF$ with $\langle \psi, \cN^m \psi \rangle \lesssim 1$ for some $m \in \bN$ large enough, and where $d_N \simeq N^{1/4}$ is chosen s.t. $\psi_N$ is normalized.

\section{A probabilistic setting}

In this last section, we formulate the convergence towards the mean field Hartree dynamics in a language more common in probability theory. To this end, we consider an $N$-particle system, described by a permutation symmetric wave function $\psi_N \in L^2 (\bR^{3N})$. We consider, moreover, a self adjoint operator $O$ acting on $L^2 (\bR^3)$. For $j=1, \dots , N$, we define $O^{(j)}$ to be the self-adjoint operator acting as $O$ on the $j$-th particle and as the identity on the other $(N-1)$ particles. Every $O^{(j)}$ can be thought of as a random variable assuming different values with different probabilities. Through the spectral theorem, the wave function $\psi_N$ determines the law of the random variables $O^{(j)}$. 

\medskip

At time $t=0$, we assume that the system is described by a factorized wave function $\psi_N = \ph^{\otimes N}$, for some $\ph \in L^2 (\bR^3)$. The operators $O^{(j)}$, $j=1,\dots , N$, define then a sequence of independent and identically distributed random variables, with a common distribution determined by  $\ph$. We obtain immediately the (weak) law of large numbers, stating that, for any $\delta >0$, 
\[ \PP_{\ph^{\otimes N}} \left( \left| \frac{1}{N} \sum_{j=1}^N \left( O^{(j)} - \langle \ph , O \ph \rangle \right) \right| \geq \delta \right) \to 0 \]
as $N \to \infty$. We also have a central limit theorem, stating that, in distribution, 
\[ \frac{1}{\sqrt{N}} \sum_{j=1}^N \left( O^{(j)} - \langle \ph , O \ph \rangle \right) \to N (0, \sigma^2) \]
where $N(0,\sigma^2)$ denotes a centered Gaussian random variable, with the variance
\[ \sigma^2 = \langle \ph , O^2 \ph \rangle - \langle \ph , O \ph \rangle^2 \, . \]

\medskip

Now, let us consider the evolution of the many body quantum system, as generated by the mean field Hamiltonian (\ref{eq:mf-ham}). The wave function describing the evolved system is given by $\psi_{N,t} = e^{-i H_N t} \ph^{\otimes N}$. For $t \not = 0$, $\psi_{N,t}$ is not factorized; hence, the random variables $O^{(j)}$ are not independent. Still, the results presented above imply that correlations are small in the limit of large $N$, at least in the sense of the reduced densities. It seems therefore natural to ask whether law of large numbers and central limit theorem are still valid at $t \not =0$.  It turns out that the convergence of the reduced density, as stated in Theorem \ref{thm:fact}, easily implies the (weak) law of large numbers. In fact, letting $\wt{O} = O - \langle \ph_t , O \ph_t \rangle$, we find
\[\begin{split} \PP_{\psi_{N,t}} \left(  \left| \frac{1}{N} \sum_{i=1}^N \wt{O}^{(i)} \right| \geq \delta \right) & \leq \frac{1}{\delta^2 N^2} \, \left\langle \psi_{N,t},  \left( \sum_{i=1}^N \wt{O}^{(i)} \right)^2 \psi_{N,t} \right\rangle \\ & = \frac{1}{\delta^2} \, \tr \, \gamma^{(2)}_{N,t} (\wt{O} \otimes \wt{O}) + \frac{1}{\delta^2 N} \, \tr \, \gamma^{(1)}_{N,t} \wt{O}^2 \\ & \to \frac{1}{\delta^2} \, \tr \, |\ph_t\rangle \langle \ph_t|^2  (\wt{O} \otimes \wt{O}) = 0 \end{split}
\]
because, by definition of $\wt{O}$, $\tr  \, |\ph_t \rangle \langle \ph_t| \wt{O} = \langle \ph_t , \wt{O} \ph_t \rangle = 0$. 

\medskip

What about the central limit theorem at $t \not = 0$? The answer to this question is given in the next theorem, which was recently proven in \cite{BKS}.
\begin{theorem}
Suppose that $V^2 (x) \leq D (1-\Delta)$ for a constant $D>0$. Let $\psi_{N,t} = e^{-iH_N t} \ph^{\otimes N}$ with $H_N$ given in (\ref{eq:mf-ham}) and for $\ph \in H^1 (\bR^3)$. Let $O$ be a bounded self-adjoint operator on $L^2 (\bR^3)$, with $\| \nabla O (1-\Delta)^{-1/2} \| < \infty$. Then, for every $t \in \bR$, with respect to the evolved wave function $\psi_{N,t}$, the random variable 
\[ \frac{1}{\sqrt{N}} \sum_{i=1}^N  \left( O^{(i)} - \langle \ph_t , O \ph_t \rangle \right)  \]
converges in distribution, as $N \to \infty$, to a centered Gaussian random variable with variance
\[ \begin{split} \sigma_t^2 = \Big[ & \, \left\langle \theta (t;0) \left(O\ph_t , \overline{O \ph_t} \right) , \theta (t;0) \left(O \ph_t , \overline{O \ph_t} \right) \right\rangle \\ &\hspace{3cm} - \left| \left\langle \theta (t;0) \left(O \ph_t , \overline{O \ph_t} \right) , \frac{1}{\sqrt{2}} \, \left(\ph, \overline{\ph} \right) \right\rangle \right|^2 \Big] \end{split} \]
where $\theta (t;0)$ is the time-dependent Bogoliubov transformation introduced in (\ref{eq:BTU}).\end{theorem}
Hence, for $t \not = 0$, the central limit theorem is still valid but the variance of the limiting Gaussian variable is changed. While on the level of the law of large numbers there is no difference between $\psi_{N,t}$ and the product $\ph_t^{\otimes N}$, this is not true for the central limit theorem; although in both cases the fluctuations are gaussians, their variance is not the same.


\frenchspacing
\begin{thebibliography}{7}

\bibitem{BKS} G. Ben Arous, K. Kirkpatrik and B. Schlein: {\sl A central limit theorem in many-body quantum dynamics.} Preprint arXiv:1111.6999. 

\bibitem{CLS} L. Chen, J.O. Lee and B. Schlein: {\sl Rate of convergence towards {H}artree dynamics.} {\it J. Statist. Phys.} {\bf 144} (2011), no. 4, 872-903.

\bibitem{EY} Erd{\H{o}}s, L.; Yau, H.-T.: Derivation
of the nonlinear {S}chr\"odinger equation from a many body {C}oulomb
system. \textit{Adv. Theor. Math. Phys.} \textbf{5} (2001), no. 6, 1169--1205.

\bibitem{GV} Ginibre, J.; Velo, G.: The classical
field limit of scattering theory for non-relativistic many-boson
systems. I and II. \textit{Commun. Math. Phys.} \textbf{66} (1979),
37--76, and \textbf{68} (1979), 45--68.

\bibitem{H} Hepp, K.: The classical limit for quantum mechanical
correlation functions. \textit{Commun. Math. Phys.} \textbf{35}
(1974), 265--277.

\bibitem{KP} Knowles, A.; Pickl, P.: Mean-field dynamics: singular potentials and rate of convergence. {\it Comm. Math. Phys.} {\bf 298} (2010), 101--139.

\bibitem{RS}
Rodnianski, I.; Schlein, B.: Quantum fluctuations and rate of convergence towards mean field dynamics. {\it Comm. Math. Phys.} {\bf 291} (2009), no. 1, 31--61.

\bibitem{S} Spohn, H.: Kinetic equations from Hamiltonian dynamics.
   \textit{Rev. Mod. Phys.} \textbf{52} (1980), no. 3, 569--615.

\end{thebibliography}
\end{document}